# Developing a Repeating Model Using the Structured Spreadsheet Modelling and Implementation Methodology

Paul Mireault
*Founder, SSMI International*
*Honorary Professor, HEC Montréal*
Paul.Mireault@SSMI.International

**ABSTRACT**

*Spreadsheets often have variables and formulas that are similar, differing only by the fact that they refer to different instances of an entity. For example, the calculation of the sales revenues of the South and East regions are Revenues South = Price*Quantity Sold South and Revenues East =Price*Quantity Sold East. In this paper, we present a conceptual modelling approach that takes advantage of these similarities and leads the spreadsheet developer to the formula Revenues = Price*Quantity. We then present simple but strict rules to implement the spreadsheet.*

## 1 INTRODUCTION

Errors in spreadsheets have caused financial losses for many companies and organizations, as illustrated by the EuSpRIG Horror Stories [EuSpRIG, 2015] web page. Panko [Panko, 2008] cites a study reporting that 95% of spreadsheets have errors.

Many authors have identified the following spreadsheet characteristics that can cause errors:

- **Far references**. A formula that references a cell that is not immediately visible and understood is harder to understand [Raffensperger, 2003].

- **Transitive references**. Formulas that reference a reference of a variable are candidates to maintenance problems. When we introduce a nuance and create a new variable, formulas may refer to one or the other nuance of the variable.

In [Mireault, 2015], we introduced the Structured Spreadsheet Modelling and Implementation methodology and illustrated it with a simple problem. The methodology is based on well-established concepts of Computer Science, Software Engineering and Information Systems. The basic idea of the methodology is to develop spreadsheets in two steps, the conceptual model first and the implementation second.

One of the important conception rules is to keep formulas as simple as they can be, avoiding having more than one mathematical operator or function in the definition of a variable. For example, the formula `Total Cost = Fixed Cost + Unit Cost * Quantity` uses two different mathematical operators, addition and multiplication, and should be replaced with `Variable Cost = Unit Cost * Quantity` and `Total Cost = Fixed Cost + Variable Cost`. Such simpler formulas would have a low complexity score according to [Hermans, et al., 2012].

[Mireault, et al., 2015] showed how a model developed with the SSMI methodology can be easily expanded to transform parameters that were entered by user into variables that are calculated from other inputs.

In this paper, we present an extension to the SSMI methodology that is used to model cases where a set of formulas and variables is repeated for different instances of an entity. For example, the calculation of the total cost is similar for all our regions, South, East and North.

## 2    MODELLING A REPEATING SUB-MODEL

We are sometimes faced with the situation where sets of variables have formulas that are similar. In such situations, we are also tempted to name the variables we create with the same prefix and differentiate them with a different suffix. For example, we might have variables named `Profit Region A`, `Profit Region B` and `Profit Region C`. If we use the straightforward modeling technique presented in [Mireault, 2015], we will end up with a model that is unwieldy and difficult to modify. But there is a way of keeping the model simple: it consists in identifying variables and formulas that are similar and grouping them in what we will call a *repeating sub-model*.

We will first illustrate the development of a model without the use of a repeating sub-model to illustrate its complexity. The reader should keep in mind that this is not the proper modeling technique. We will then illustrate the proper use of the repeating sub-model and show how it simplifies the Formula Diagram and the Formula List.

Let's consider a small example with Marco's Widgets. Marco sells his widgets in three different regions: South, East and North. He wants a spreadsheet that will show him the profit per region as well as the total profit. To allocate the demand per region, he tells you that the demand has traditionally been 48%, 23% and 29% respectively for the South, East and North regions. Marco uses a `Unit Cost` composed of a `Manufacturing Cost` and a `Delivery Cost`. The `Manufacturing Cost` does not depend on the region and is equal to 120$. The `Delivery Cost` is equal to 50$, 80$ and 60$ respectively for the South, East and North regions. To calculate each region's profit, the `Fixed Cost` will be allocated to each region with the same distribution as the demand. We now have all the information needed to design the model.

### 2.1    Building the model for the South region

*Figure 1 presents the Formula Diagram for the South region only. The corresponding Formula List is shown in*

Table 1.

Now, if we continue on with the East region, we will notice that the variables and the formulas are similar. The only difference is that the variables will use the `East` suffix instead of `South`.

It will also be the same with the North-region and the `North` suffix. We would end up with a model that looks like Figure 2.

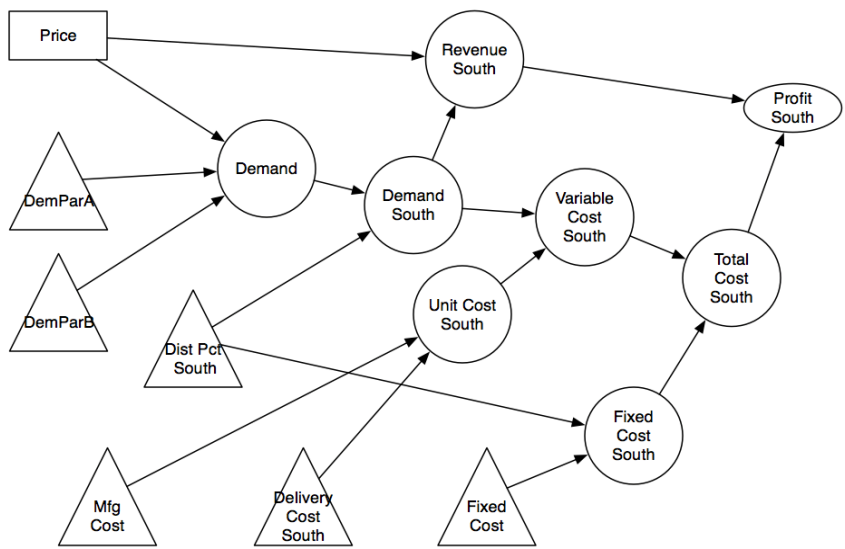

*Figure 1 - Formula Diagram for the South region*

| Variable | Description | Type | Definition |
|---|---|---|---|
| `Price` | Average price of widgets | Input | |
| `Profit South` | Profit of the South region | Output | `= Revenue South - Total Cost South` |
| `DemParA` | First Demand function parameter | Parameter | `367,000` |
| `DemParB` | Second Demand function parameter | Parameter | `1.0009` |
| `Fixed Cost` | Fixed cost of manufacturing the widgets | Parameter | `2,500,000$` |
| `Mfg Cost` | Cost of manufacturing one widget | Parameter | `120$` |
| `Dist South` | Proportion of the Demand sold in the South region | Parameter | `49%` |
| `Delivery Cost South` | Cost of delivery of widgets in the South region | Parameter | `50$` |
| `Demand` | Demand of widgets, formula given by the market research specialist | Calculated | `= DemParA * DemParB^-Price` |
| `Demand South` | Portion of the Demand sold in the South region | Calculated | `= Demand * Dist South` |
| `Total Cost South` | Total Cost of selling widgets in the South region | Calculated | `= Fixed Cost South + Variable Cost South` |
| `Fixed Cost South` | Portion of the Fixed cost allocated to the South region | Calculated | `= Fixed Cost * Dist South` |
| `Variable Cost South` | Variable Cost of the widgets sold in the South region | Calculated | `= Demand South * Unit Cost South` |
| `Unit Cost South` | Unit cost of one widget in the South region | Calculated | `= Mfg Cost + Delivery Cost South` |
| `Revenue South` | Revenue of the South region | Calculated | `= Demand South * Price` |

*Table 1 - Formula List of the South region model*

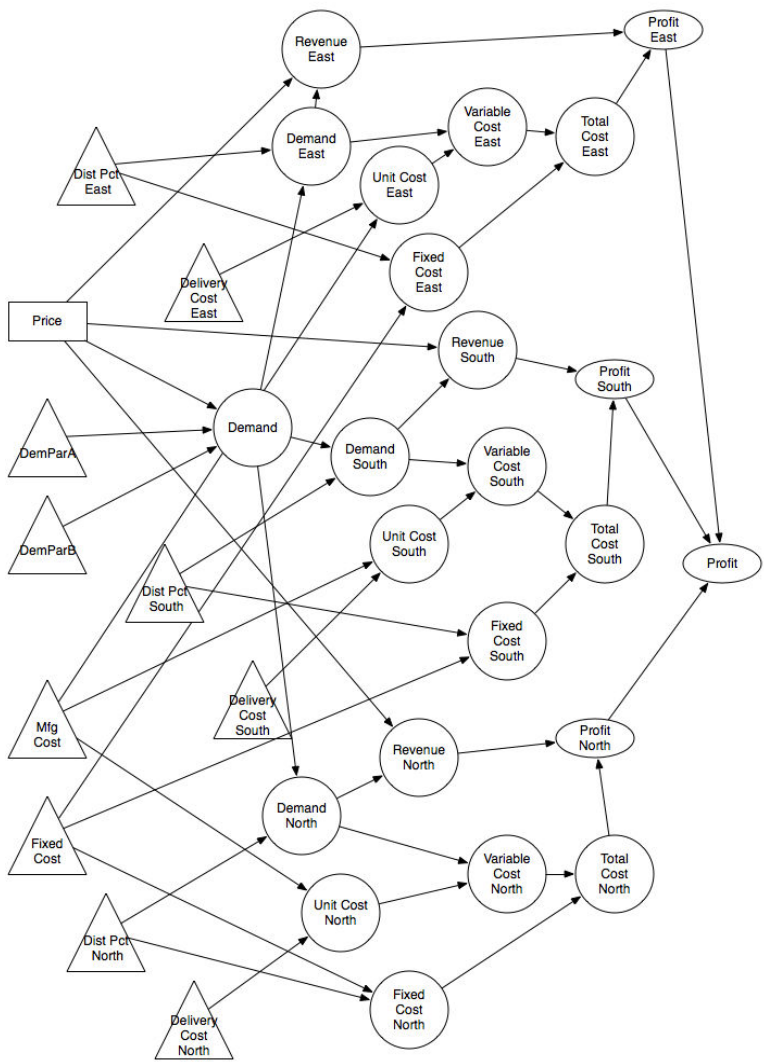

*Figure 2 - Formula Diagram for three regions*

## 2.2 Creating a repeating sub-model

The model of Figure 2 has obvious shortcomings. The most important one is that it does not scale well. Imagine how it would look like if we had to expand it to cover more regions, like provinces or states in many countries. Canada has 10 provinces, India has 28 or more states, the USA has 50 states, France has 100 departments; expanding the model to cover the divisions of many countries is practically infeasible. Another important shortcoming is that any modification, like adding variables, has to be replicated many times. This increases the risk of introducing errors, which is something we want to avoid.

The key to the repeating sub-model lies in the variables that have a suffix. Instead of having one *copy* of a variable for each region, we will use *one* variable representing any region. The variables `Delivery Cost South`, `Delivery Cost East` and `DeliveryCostNorth` will be replaced by `Delivery Cost`. Thus, `Delivery Cost` is now a variable with multiple values: it is the set of 3 values {50, 80, 60}.

When the same variable name would appear more than once, we need to use an adjective to identify its role. For instance, we have two variables that represent the demand: the variable `Demand` that represents a single value and the variable `Demand` that represents multiple values will be renamed `Total Demand` and `Regional Demand` respectively.

We will apply the same treatment to the other variables. The repeating variables are `Distribution`, `Regional Demand`, `Delivery Cost`, `Unit Cost`, `Revenue`, `VariableCost`, `Regional Fixed Cost`, `Total Cost` and `Profit`.

We represent the *repeating sub-model* with a box with a dash border in the Formula Diagram. We write the name of the repeating entity in the top right corner of the box. Now, we place all the variables that have multiple values, depending on the Region, inside the sub-model box and all the other variables outside the box. This is illustrated in Figure 3.

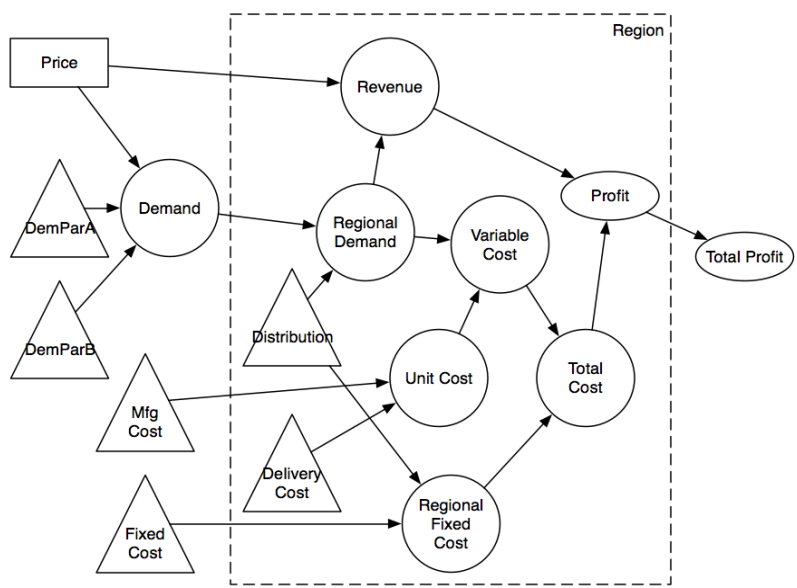

*Figure 3 - Formula Diagram with the region sub-model*

All the variables outside the box represent single values, and all the variables inside the box represent multiple values.

This method of representing a *repeating sub-model* takes care of the two major shortcomings we mentioned earlier. First, the model is now scalable as it does not get more complicated with an increase in the number of regions. In fact, the model does not change at all when we increase the number of regions. Naturally, the spreadsheet implementation will be bigger with more regions, but we will show a straightforward way of expanding the spreadsheet implementation to accommodate more regions.

Second, modifying the model is simplified by the fact that parts of the model are not repeated many times. Any change needs only to be made once, no matter how many regions there are.

We calculate the `Total Profit` from the `Profit` of each region. Since it is inside the repeating sub-model box, the variable `Profit` represents a set of values, and since it is outside the box, the variable `Total Profit` represents a single value. Thus, the function that calculates `Total Profit` must be a function that takes a set of values and returns a single value. There are a few such functions, called **aggregate functions**, and you are already familiar with some them: SUM,

AVERAGE, MINIMUM, MAXIMUM, VARIANCE, and STANDARD DEVIATION. There are others that are more specialized, like NPV (Net Present Value) and IRR (Internal Rate of Return) often used in Finance and Accounting models.

In our case, we will use the formula `Total Profit = SUM(Profit)`. We add the following entry in our Table of Formulas:

| Variable | Description | Type | Definition |
| --- | --- | --- | --- |
| `Price` | Average price of widgets | Input | |
| `Profit` | Profit of each region | Output, repeating | `= Revenue - Total Cost` |
| `DemParA` | First Demand function parameter | Parameter | `367,000` |
| `DemParB` | Second Demand function parameter | Parameter | `1.0009` |
| `Fixed Cost` | Fixed cost of manufacturing the widgets | Parameter | `2,500,000$` |
| `Mfg Cost` | Cost of manufacturing one widget | Parameter | `120$` |
| `Distribution` | Proportion of the Demand sold in each region | Parameter, repeating | `48%, 23%, 29%` |
| `Delivery Cost` | Cost of delivery of widgets in each region | Parameter, repeating | `50$, 80$, 60$` |
| `Total Demand` | Demand of widgets, formula given by the market research specialist | Calculated | `= DemParA * DemParB^-Price` |
| `Regional Demand` | Portion of the Demand sold in each region | Calculated, repeating | `= Total Demand * Distribution` |
| `Total Cost` | Total Cost of selling widgets in each region | Calculated, repeating | `= Regional Fixed Cost + Variable Cost` |
| `Regional Fixed Cost` | Portion of the Fixed cost allocated to each region | Calculated, repeating | `= Fixed Cost * Distribution` |
| `Variable Cost` | Variable Cost of the widgets sold in each region | Calculated, repeating | `= Regional Demand * Unit Cost` |
| `Unit Cost` | Unit cost of one widget in each region | Calculated, repeating | `= Mfg Cost + Delivery Cost` |
| `Revenue` | Revenue of each region | Calculated, repeating | `= Regional Demand * Price` |
| `Total Profit` | Total profit of all regions | Output | `= SUM(Profit)` |

*Table 2 - Formula List with the repeating sub-model*

## 3   THE REPEATING SUB-MODEL IMPLEMENTATION

Like the implementation of the simple model shown in [Mireault, 2015], the implementation of the repeating sub-model follows precise rules. These rules have been empirically designed to reduce the possibility of making errors during the initial implementation of the spreadsheet as well as during its maintenance. All definition formulas are identified with a bold font and a proper number format (currency or percentage) is used where applicable.

We will need five worksheets. As described in [Mireault, 2015], we have the `Interface` sheet that will be used by the spreadsheet's user. We also need the

`Parameters` and the `Model` sheets for all the variables and constants that are **outside** the repeating sub-model box. We will also use two sheets for the repeating parameters and the repeating variables defined **inside** the repeating sub-model box. We will name them `Parameters-Regions` and `Regions`.

**The interface sheet**

In the `Interface` sheet (Figure 4), we put the input variables with their reasonable starting value. We then *name* the cells containing the input variables. Finally, we prepare the area where we will put the references to the output variables. In the case where some output variables are from the repeating sub-model, we will also reference the repeating entity.

We will come back to complete Output Variables section of the `Interface` sheet after we have finished with the model.

*Figure 4 - Step 1: Naming the sheets and setting up the Interface*

**Step 2: The parameters sheets**

We define the single value parameters by putting the labels in column A and the values in column B. As illustrated in Figure 5, we name the single value parameters by selecting the labels and the values and clicking on the *Create from Selection* button of the *Formulas* ribbon.

*Figure 5 - Step 2: Creating names for single value parameters*

The repeating entity is defined in the `Parameters-Regions` sheet with the column labels we will use to identify its instances. In column A, we first write the name of the repeating entity, `Region` in our case, and the names of the parameters appearing inside the repeating sub-model box. Then, starting in Column B, we write the values of the entity and of the parameters, as illustrated in Figure 6.

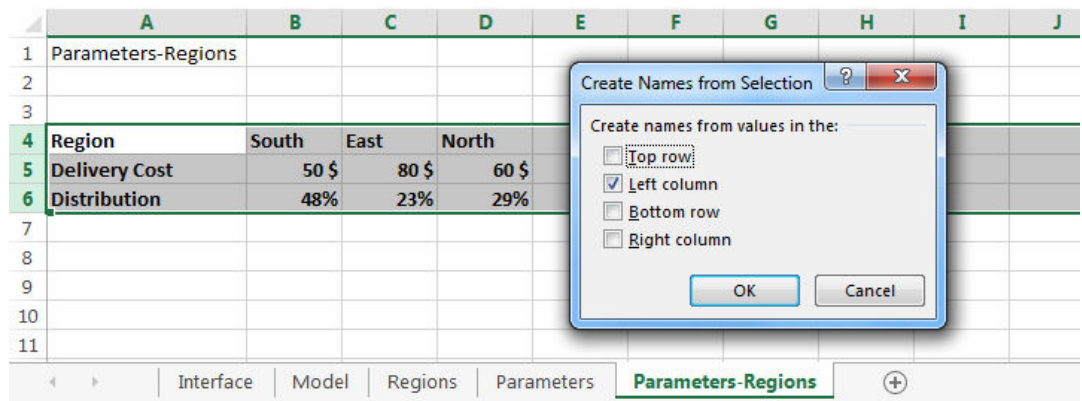

*Figure 6- Step 2: Parameters of the repeating entity*

Finally, we name the multiple value parameters by first selecting the **entire rows** consisting of the labels all the values and all the blank cells following them, as shown in Figure 7.Naming the whole row like this will allow us to easily add new regions.

*Figure 7- Step 2: Naming the repeating entity parameters*

**Step 3: The Model sheet**

*In the Model sheet, we need to put the definition formulas of all the calculated variables that are outside of the repeating sub-model box. In our case, there are only two variables: `TotalDemand` and `Total Profit`.*

Figure 8 shows the block structure with the defining variables above the line and the defined variable below. It also shows that cell B2 has been named `Total Demand`. Since `Total Profit` is calculated using a variable defined in the repeating sub-model, we will defer defining it until we have completed the repeating sub-model.

*Figure 8 - Step 3: The Model sheet*

**Step 4: The repeating sub-model Sheet**

Implementing a repeating sub-model is done in two phases. The first phase consists of implementing the model **for only one instance of the repeating entity** by following the usual block structure and variable naming operations. The second phase consists of **one** copy operation where the model we implemented for the first instance of the entity is copied for all the other instances.

We start by identifying the region corresponding to the first column of the model. As shown in Figure 9, we put the label `Region` in cell A3 and the formula `=Region` in B3. It is important to start the model in the same column we used in the Parameters sheet; otherwise the *names* will not work properly.

*Figure 9–Step 4: Setting up the repeating entity*

We then develop the model with the block structure described in [Mireault, 2015] with the exception that we use the label of column A to name the whole row instead of the cell on its right as shown in Figure 10.

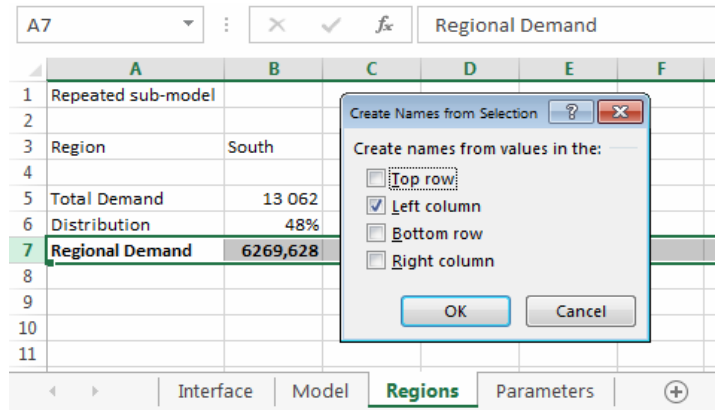

*Figure 10– Step 4: Naming the whole row of a variable's definition*

*Figure 11 – Step 4: The completed model for one region*

*Figure 12 – Step 4: The Formula View of the completed model*

Figure 11 shows the block structure for all the variables of the repeating sub-model. We can see in its corresponding Formula View (Figure 12) the structure of each *definition* block: the top part contains named references to the variables used in the calculation, and the bottom part is the actual definition formula. By using only the cells directly above, the definition formula is easy to audit.

Once the model for one region is completed we will copy it to the right for the other regions. We start by selecting column B by clicking on its column letter as illustrated in Figure 13. We then drag the *copy handle* two columns to the right. The final result is shown in Figure 14.

*Figure 13- Step 4: Selecting the entire model of one region*

*Figure 14 - Step 4: Copying the full model*

Finally, we return to the Model sheet to implement the definition of `Total Profit`. Since it uses a repeating variable in its calculation we will implement it with a variation of the block structure.

First, we set up the usual block structure, with the references in the top part, as shown in Figure 15. We also include references to the repeating entity as a visual reference.

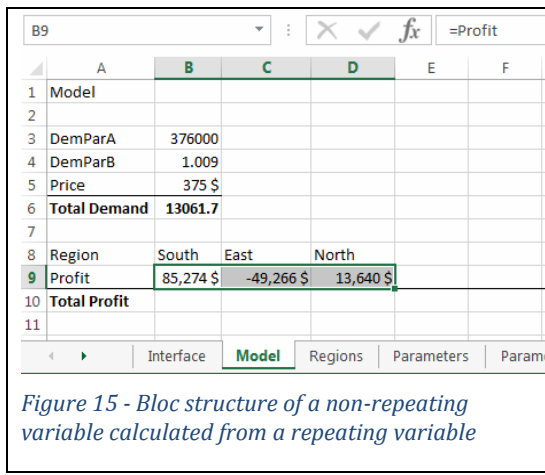
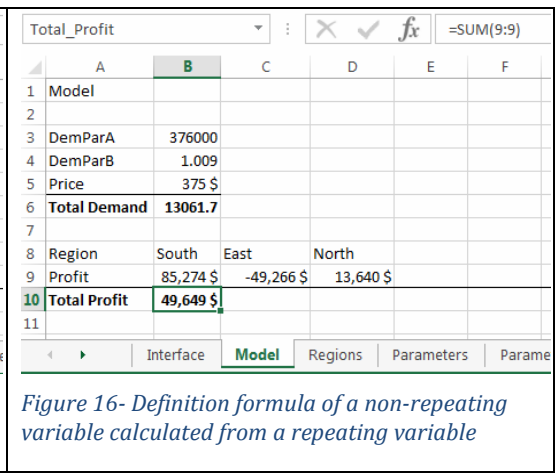

*Figure 15 - Bloc structure of a non-repeating variable calculated from a repeating variable*

*Figure 16 - Definition formula of a non-repeating variable calculated from a repeating variable*

Next, we write the formula in cell B10 as the SUM of the whole row above it. As shown in Figure 16, we name the single cell B10, not the whole row 10, but we put the top border on the whole row as a visual indication that cell B10 uses the whole row above it.

**Step 5: Finishing the Interface sheet**

The final step consists of putting the references to the output variables in the Interface sheet. In column B we write the reference formulas for the entity, `Region`, and the output variables, `Profit`, as illustrated in Figure 17.

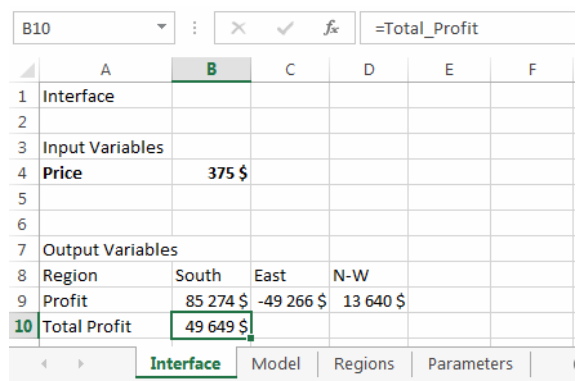

*Figure 17 – Step 5: Finishing the Interface sheet*

### 4    CONCLUSION

We have shown how to model a problem involving a repeating entity in a way that creates a simple and elegant sub-model.

By separating the creative task of *building the conceptual model* from the mechanical task of *implementing the physical model*, we expect to reduce *logical errors* due to the constant interruptions of the brain's creative activities.

By reducing and by tightly constraining the *copy* operations, we expect to reduce the number of errors that are due to manipulations.

Further research can demonstrate whether the use of the Structured Modelling and Implement methodology with repeating sub-model does indeed affect the probability of making errors in a spreadsheet.